# Hole- and electron-injection driven phase transitions in transition metal dichalcogenides and beyond: A unified understanding

Xiao-Huan Lv, Meng-Qi Wu, Yin-Ti Ren, Rui-Ning Wang,* Hu Zhang, Chen-Dong Jin, Ru-Qian Lian, Peng-Lai Gong, Xing-Qiang Shi,* and Jiang-Long Wang*

Key Laboratory of Optic-Electronic Information and Materials of Hebei Province, Institute of Life Science and Green Development, College of Physics Science and Technology, Hebei University, Baoding 071002, P. R. China

*E-mails: rnwang@hbu.edu.cn; shixq20hbu@hbu.edu.cn; jlwang@hbu.edu.cn

**Abstract:** The phase transitions among polymorphic two-dimensional (2D) transition metal dichalcogenides (TMDs) have attracted increasing attention for their potential in enabling distinct functionalities in the same material for making integrated devices. Electron-injection to TMDs has been proved to be a feasible way to drive structural phase transition from the semiconducting H-phase to the semimetal dT-phase. In this contribution, based on density-functional theory (DFT) calculations, firstly we demonstrate that hole-injection drives the transition of the H-phase more efficiently to the metallic T-phase than to the semimetallic dT-phase for group VI-B TMDs ($MoS_2$, $WS_2$, and $MoSe_2$, etc.). The origin can be attributed to the smaller work function of the T-phase than that of the dT-phase. Our work function analysis can distinguish the T and dT phases quantitatively while it is challenging for the commonly used crystal field splitting analysis. In addition, our analysis provides a unified understanding for both hole- and electron-injection induced phase transitions for 2D materials beyond TMDs, such as the newly synthesized $MoSi_2N_4$ family. Moreover, the hole-driven T-phase transition mechanism can explain the recent experiment of $WS_2$ phase transition by hole-doping with yttrium (Y) atoms. Using 1/3 Y-doped $WS_2$ and $MoSe_2$ as examples, we show that the Mo and W valency increases to 5+. These above findings open up an avenue to obtain the metallic T-phase, which expands the possible stable phases of 2D materials.

# I. INTRODUCTION

Two-dimensional transition metal dichalcogenides (TMDs) have attracted increasing attention in fundamental research and device applications for their unique structural, physical and chemical properties[1-4]. The group VI-B TMD monolayers, $MX_2$ (M = W, Mo; X = S, Se, Te), have several polymorphs such as the H, T and dT phases[5-7]. In the H structure, the X-M-X atomic layers form AbA stacking (Fig. 1a); while in the T structure, the three atomic layers form AbC stacking (Fig. 1b). The T-phase can be obtained by alkali atom intercalation, and may coexist with the other phases[8]. The isolated T-phase structure is generally unstable and undergoes spontaneous lattice distortion along the $x$ direction (as denoted in Fig. 1c), forming the dT-phase with a doubled ($\sqrt{3} \times 1$) rectangular cell, where the distorted M atoms form one-dimensional zigzag chains in the $y$ direction as indicated by the dashed lines in Fig. 1c[9,10]. For the H-phase TMDs, many have semiconducting properties with spin-orbit coupling and exciton effects for various applications such as field effect transistors[11,12], magnetic tunnel junctions[13], valleytronics[14] and optoelectronics[15-17]. For the T-phase and its distorted form (the dT-phase), large magneto resistance, interesting quantum spin Hall effect[18] and higher catalytic activity[19] have been reported. Recently, the phase engineering and phase transition between polymorphic 2D TMDs have attracted increasing attention for the potential of enabling distinct functionalities in the same material. For example, the integration of H and dT phases in $MoTe_2$ thin film by laser-induced transition have been shown to be an Ohmic homojunction[20].

Different methods have been reported to induce phase transition from the H- to the dT- or T-phase, such as the intercalation[21] of alkali metal ions Li, Na, and K[22], injection of electrons[23,24], applying stresses[25,26], changing temperature[6,10], laser and focused electron beams[20,27], and doping with transition metal elements[28-32]. A recent theoretical work has demonstrated that electron-injection is a feasible mean to drive structural transition from the semiconducting H-phase to semimetal dT-phase[23], and, recently the hole-injection induced transition to metallic T-phase (not dT) has been realized in the experiment[33]. However, the mechanism of hole-injection induced T-phase transition of TMDs and a unified understanding for both hole- and electron-driven phase transitions in TMDs and beyond remain lack. The possible reason is that it is challenging for the commonly used crystal field splitting analysis to distinguish quantitatively between the T- and dT-phase. New analysis methods are needed.

In the current work, we conducted a systematic density functional theory study to provide a unified understanding to both hole- and electron-injection driven phase transitions of group VI-B TMDs and beyond, and then realized the proposed hole-injection driven phase transition in real systems. We found that hole-injection drives the H-phase transition more efficiently to the metallic T-phase than to the semimetallic dT-phase, and then revealed the phase transition mechanism through an analysis of the different work functions between the different phases. Our simple work function analysis can distinguish the T and dT phases quantitatively. Finally, inspired by the recent experiment[33], we used yttrium (Y)-doped $MX_2$ to demonstrate the realization of hole-injection driven phase transition in real systems, in which the valency of M atoms increases to 5+ under 1/3 Y concentration.

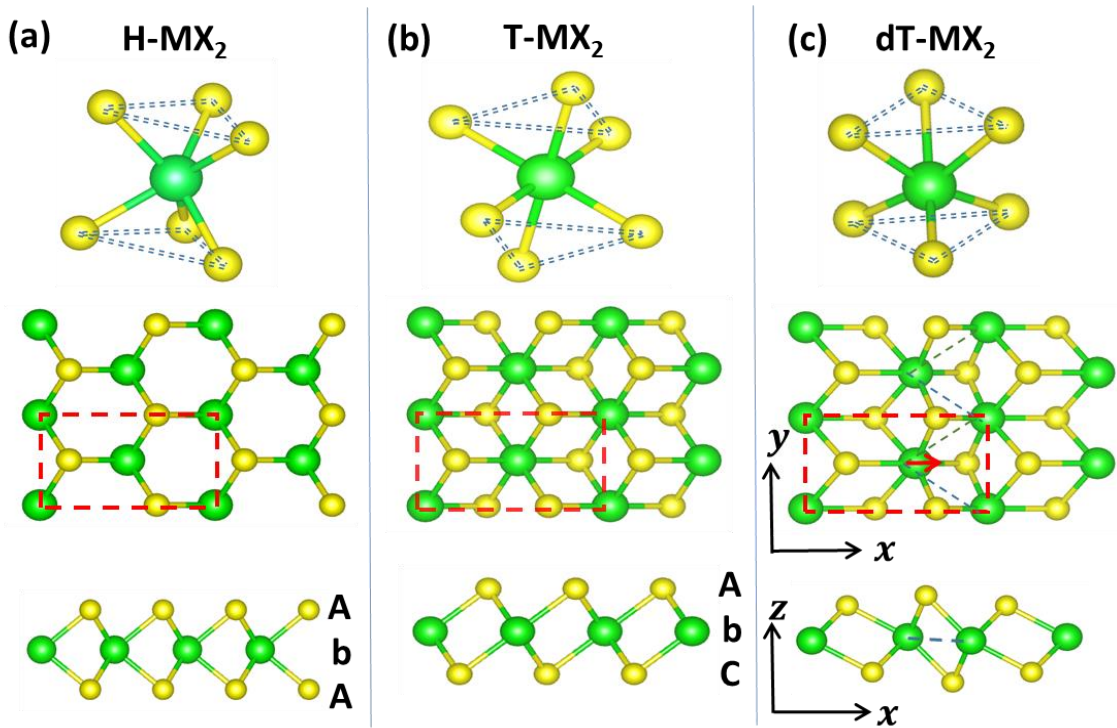

FIG. 1. Atomistic structures of monolayer transition metal dichalcogenides, $MX_2$ (M = Mo, W; X = S, Se, Te), in different phases. (a) 1H-$MX_2$ in AbA stacking; (b) 1T-$MX_2$ in AbC stacking; and (c) distorted 1T-$MX_2$ (dT-$MX_2$), where the distorted M atoms form one-dimensional zigzag chains as indicated by the dashed lines along the $y$-direction. Unit cells are indicated by red rectangles; the cell sizes in the H- and T-structure are doubled to compare with the doubled ($\sqrt{3} \times 1$) rectangular cell in the dT-structure.

## II. CALCULATION METHODS

Spin-polarized density functional theory calculations were performed using the Vienna *ab initio* Simulation Package (VASP)[34,35] The projector augmented wave (PAW) potentials was adopted to describe the core electrons[36]. The valence electrons were described by plane waves with a cut-off kinetic energy of 400 eV. The electronic exchange-correlation energy was treated by the generalized-gradient approximation of Perdew-Burke-Ernzerhof (GGA-PBE)[37] form. A vacuum layer of ~20 Å along the *z*-direction was added to avoid the spurious interaction between neighboring slabs. The *K*-point sampling in the Brillouin zone adopts the Monkhorst-Pack scheme with a grid density of 11×11×1 and 7×11×1 for $1 \times 1$ and $\sqrt{3} \times 1$ cells, respectively. A similar *K*-point density was used for other calculations. The convergence criteria of energy and forces acting on each atom were $10^{-5}$ eV and 0.01 eV Å$^{-1}$, respectively. For bonding analysis, the LOBSTER package was used[38].

## III. RESULTS AND DISCUSSIONS

The lattice parameters and electronic band structures of group VI-B TMD monolayers, $MX_2$ (M = Mo, W; X = S, Se and Te), are summarized in Table S1 and Fig. S1 in the Supporting Information. Fig. 2 compares the relative energetic stabilities among the H, T and dT phases, which is the starting point to study the phase transitions among them induced by hole- and electron-injections. The energy differences among the three phases change monotonously with the increase of the period of X atom from S to Se to Te for both $MoX_2$ and $WX_2$. In detail, the energy difference between the T (or dT)-phase and the H-phase, E(T)-E(H) or E(dT)-E(H), decreases as the period of the X atom increases; while the energy difference between T and dT, E(T)-E(dT), increases with the increase of the X atom period. For $MoTe_2$, the H and dT phases can coexist in experiment[2], while the stable structure of $WTe_2$ in experiment is the dT-phase[10,39,40].

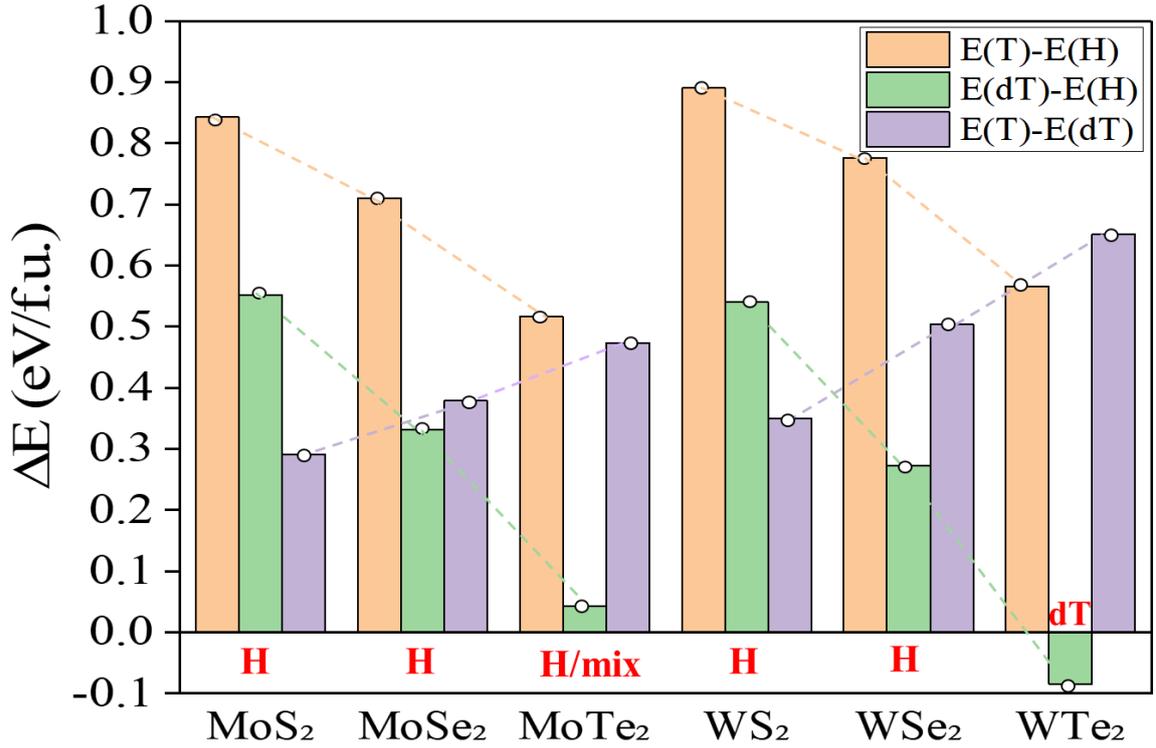

FIG. 2. Energy differences are in units of eV per formula unit (eV/f.u.) among the three phases of group VI-B TMD monolayers. The energy differences change monotonously with the increase of the period of X atom from S to Se to Te for both $MoX_2$ and $WX_2$ (see the dashed lines guide the eye). The labeled "H", "H/mix", and "dT" in the lower part of the figure denotes the stable structure in experiment is the H-phase, the mixed H and dT phases, and the dT-phase, respectively.

### A. Trend of phase transitions

It is well known that the band edge of $MX_2$ is mainly contribute by the $d$-orbitals of the M atoms[1,41]. An analysis of the energy level splitting of M $d$-orbitals is helpful for exploring the mechanism of phase transitions. The crystal field splitting and electron filling of the $d$ shells of group VI-B transition metal atoms in different phases are sketched in Fig. 3a. For the H-phase (Mo in trigonal prismatic coordination), the $d$ energy levels split into a singlet $a_1$ (low-energy $d_{z^2}$), a doublet $e$ (medium-energy $d_{xy}/d_{x^2-y^2}$), and a doublet $e'$ (high-energy $d_{yz}/d_{xz}$)[42,43]. While for the T-phase (octahedral coordination), the $d$ orbitals split into a low-energy triplet $t_{2g}$ and a high-energy doublet $e_g$. For the dT-phase, Mo in a distorted octahedral coordination and also interact (weakly) with two neighboring Mo atoms (the bottom panel of Fig. 3a), so that there is a further splitting within the $t_{2g}$ and $e_g$ levels. Partial orbital occupation of $t_{2g}$ leads not only to the metallic (semimetallic) properties of the T(dT)-phase but also to the structural instability, which explains why the H-phase is more ubiquitous under environmental conditions for most MX$_2$. While monolayer MX$_2$ cannot transform its phase spontaneously under neutral conditions, recent studies have shown that doping electrons (holes) to two dimensional MX$_2$ may transform it from the H-phase to the dT(T)-phase.[23,33]

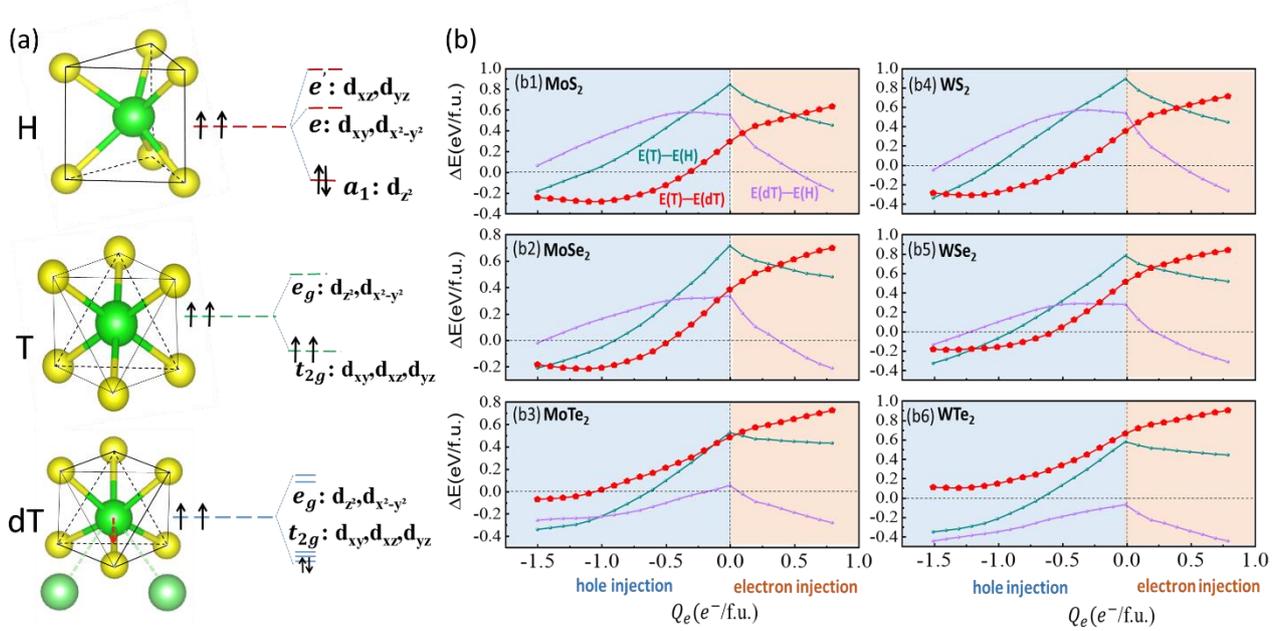

FIG. 3. (a) Coordination, $d$-levels splitting and electron filling of the transition metal atom of group VI-B TMDs in the H-,T- and dT-phase. (b) Energy differences (ΔE) as a function of electron and hole injections between the three phases [E(T)-E(dT), E(T)-E(H) and E(dT)-E(H)]. For hole injection, the H-phase transition more efficiently to the metallic T-phase than the semimetallic dT-phase, especially for MoS$_2$, WS$_2$, MoSe$_2$, and WSe$_2$.

Fig. 3b shows the energy differences among the three phases [E(T)-E(dT), E(T)-E(H) and E(dT)-E(H)] of group VI-B MX$_2$ as a function of electron or hole injection. For electron injection, the E(dT)-E(H) results in Figs. 3b1-b3 reproduces that of the literature[23] well, indicating that our calculation are reliable. Interestingly, Fig. 3b demonstrates that for hole injection, the H-phase transition more efficiently to the metallic T-phase than to the semimetallic dT-phase, especially for MoS$_2$, WS$_2$, MoSe$_2$ and WSe$_2$. The $d$-levels splitting and electron filling in Fig. 3a cannot explain this, since it is difficult to distinguish quantitatively between the T and dT phases from crystal field splitting. We then quantify the degree of difficulty of removing (hole injection) or adding electrons (electron injection) from the three phases by inspecting their ionization energy (IE) or electron affinity energy (EA) for semiconducting H-phase, and work function (WF) for (semi)metallic T and dT phases in Fig. 4 and Fig. S2.

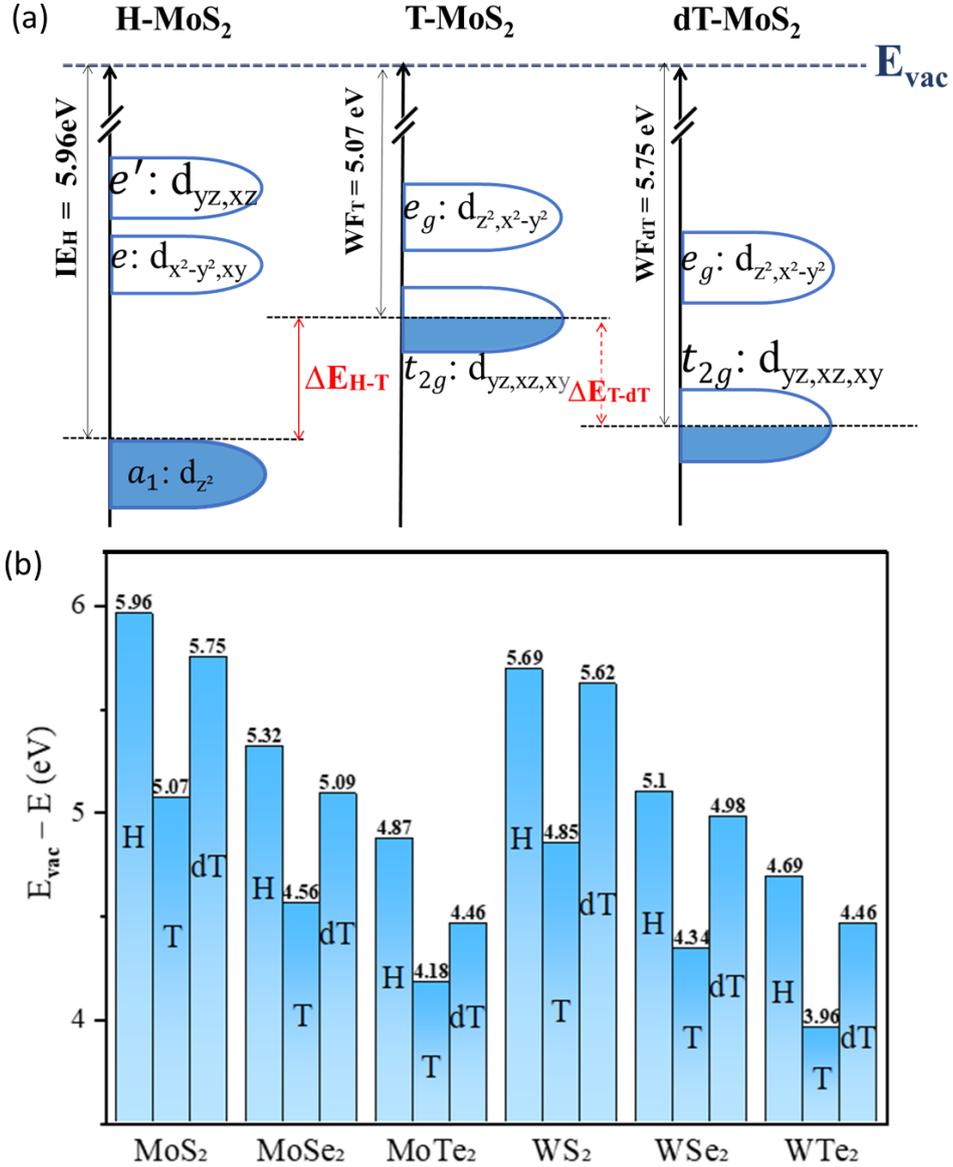

FIG. 4. Ionization energy (IE) of the H-phase and work-function (WF) of the T- and dT-phase of $MoS_2$ (a) and all group VI-B TMDs (b). It can be seen from (a) that the phase change from H to T is preferred when holes are injected (removing electrons). $E_{vac}$ is the vacuum level.

Taking $MoS_2$ as an example, Fig. 4a shows that hole-injection (removing electrons) from the H-phase is the most difficult (IE = 5.96 eV), and it is much easier to remove electron from the T-phase

(WF = 5.07 eV) than from the dT-phase (WF = 5.75 eV). Fig. 4b shows that all group VI-B $MX_2$ in the three phases show the same trend as that of $MoS_2$, showing that the T-phase is the easiest to remove electrons (hole injection). This explains why in Fig. 3b, the H-phase transition more efficiently to the metallic T-phase than the semimetallic dT-phase for hole injection. Note that, our above analysis can also apply to the H-to-dT phase transition for electron injection, by comparing the EA of the H-phase and the work function of the T and dT phases (Supporting information Fig. S2). So, our method from IE (or EA) and WF analysis has the advantages of quantify the differences between the T and dT phases, and provides a unified understanding to both the electron-induced H-to-dT phase transition and the hole-induced H-to-T phase transition.

Above demonstrates that electron (hole) injection leads to phase transition from the H-phase to the dT(T)-phase, which should not limited to TMDs, but also sheds light on the phase transition of the newly synthesized $MSi_2N_4$ family (M is a transitional metal)[44,45]. Note that due to the confinement of the outer Si-N layers, only the H and T phases are possible and the dT-phase may not exist in $MSi_2N_4$[46]. Ding *et al.* reported that the H-phase $MSi_2N_4$ are only stable for groups V-B and VI-B transition-metal systems, while for the early (groups III-B and IV-B) and late (group VIII-B) transition-metal systems, the T-phase structures are more stable or only the T-phase are stable[47]. This indicate that our conclusion can be applicable to systems beyond TMDs.

### B. Heteroatom-doped TMDs

Though it seems that more holes are needed for the T-phase transition than the electrons needed for the dT-phase transition (Fig. 3b), the hole-injection induced T-phase transition has been realized in the experiment[33]. A recent combined experimental and theoretical study by Du *et al.* has shown that the energy difference between the H-phase and T-phase of $WS_2$ decreases as the yttrium (Y) substitutional doping concentration increase[33]. They found that for the Y concentration of 1/4 the energy difference between T-phase and H-phase reduced to about 0.2 eV, and, with Y, P co-doping the stable phase can be the T-phase under 1/4 Y concentration for the P atom adsorption on top of Y. However, the dT-phase was not considered in their work.

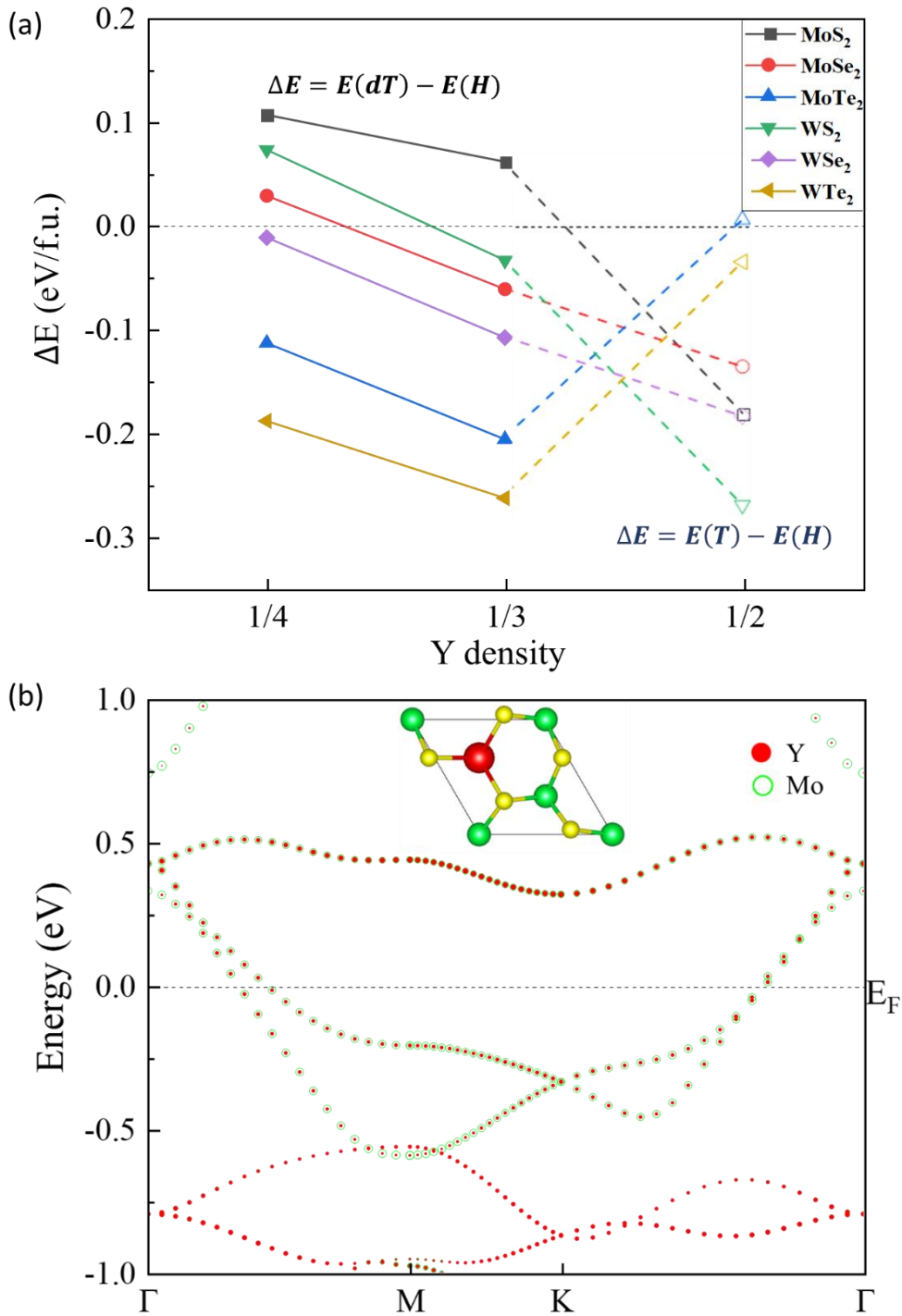

FIG. 5. (a) Energy differences between the H-phase and the dT- or T-phase of group VI-B MX$_2$ as a function of yttrium (Y) density. For the 1/4 and 1/3 Y concentrations, the dT-phase is more stable than the T-phase and E(dT)-E(H) is shown; while for 1/2 concentration, the T-phase is more stable than the dT-phase and E(T)-E(H) is shown. (b) Projected band structure for 1/3 Y-doped H-MoSe$_2$. The red and green circles denote the projected band of Y and Mo atoms, respectively.

To consider the possible dT- or T-phase transition from the H-phase with Y doping/alloying, MX$_2$ with 1/4, 1/3, and 1/2 Y concentrations are considered within $\sqrt{3} \times 2$, $\sqrt{3} \times 3$, and $\sqrt{3} \times 1$ rectangular cells, respectively. We find that Y doping can make the T-phase stable under 1/2 Y concentration (Fig. 5a). Note that in literature[33], Y, P co-doping can obtain the T-phase under 1/4 Y concentration, here we show that Y only may obtain the T-phase with a higher Y concentration. For each Y concentration, all nonequivalent Y substitutional doping patterns are considered, and Fig. 5a shows the lowest energy results. For the 1/4 and 1/3 Y doping concentrations, the dT-phase is more stable than the T-phase and the energy difference of E(dT)-E(H) is shown; while for 1/2 concentration, the T-phase is more stable than the dT-phase and E(T)-E(H) is shown. It is important to note that, under 1/2 Y concentration, the dT-phase cannot exist and relaxed to the T-phase during structural relaxation.

It is interesting to note that in the Y doped systems, a phase transition from H to dT to T occurs (Fig. 5a), the origin (not only hole-doping effect) is beyond the scope of the current work and deserve more detailed study. In addition to complex phase transition, we find that Y-doping may induce the valency change of transition metal Mo and W[48]. For example, for the band structure of H-MoSe$_2$ with 1/3 Y doping, the energy band of Mo and Y are degenerate and both cross the Fermi level (Fig. 5b). According to our previous study[49], this means that the Mo valency is increased from 4+ to 5+ and the Y valency is 2+, which also confirms that Y doping injects hole into the MoSe$_2$ system. The same picture applies for Y-doped WS$_2$ (Fig. S4).

Crystal orbital Hamilton population ($COHP$) bonding analysis and its generalized version, the density of energy ($DOE$) function analysis[38,50], are further performed to understand the phase transitions with increasing Y doping concentration. Taking $(Y_xM_{1-x})Se_2$ ($x$ = 1/3, 1/2) as examples, the $COHP$ curves shows the bonding and antibonding contributions of Y-Se and Mo-Se bonds. Fig. S5 in the Supporting Information shows the $COHP$ curve at different Y concentrations of $(Y_xM_{1-x})Se_2$. For the H-phase, Y doping induces a large number of anti-bonding states around -5 eV that make the doped system unstable, which is the possible origin for the phase transition. However, for the T-phase, anti-bonding states around -5 eV also appears, although the anti-bonding is weakened near Fermi energy in the doped systems than the undoped one. So, it is difficult to quantify from the $COHP$ map. To better quantify the bonding of $(Y_xM_{1-x})Se_2$ in the H and T/dT phases, we adopt the density of energy ($DOE$) function[38], which may be seen as the generalized $COHP$:

$$DOE(E) = \sum_k \sum_A \sum_{\substack{\mu \\ \mu \in A}} \sum_B \sum_{\substack{\nu \\ \nu \in B}} P_{\mu\nu}(E,k) H_{\mu\nu}(k) \qquad (1)$$

The $DOE$ comprises all elements of the density-of-states matrix $P_{\mu\nu}$ and the Hamiltonian matrix $H\mu\nu$, which are made up from the atomic orbitals $\mu$ and $\nu$. It sums up all offsite (A-B interatomic) and on-site (intra-atomic) contributions. The entire band energy is given by

$$E_{band} = \int_{-\infty}^{\varepsilon_F} DOE(E) dE \qquad (2)$$

Fig. 6 shows the $DOEs$ and the corresponding band energies (in blue) for the cases with H-to-dT/T phase transitions. The $E_{band}$ value for the dT- or T-phase is always greater than that of the H-phase.

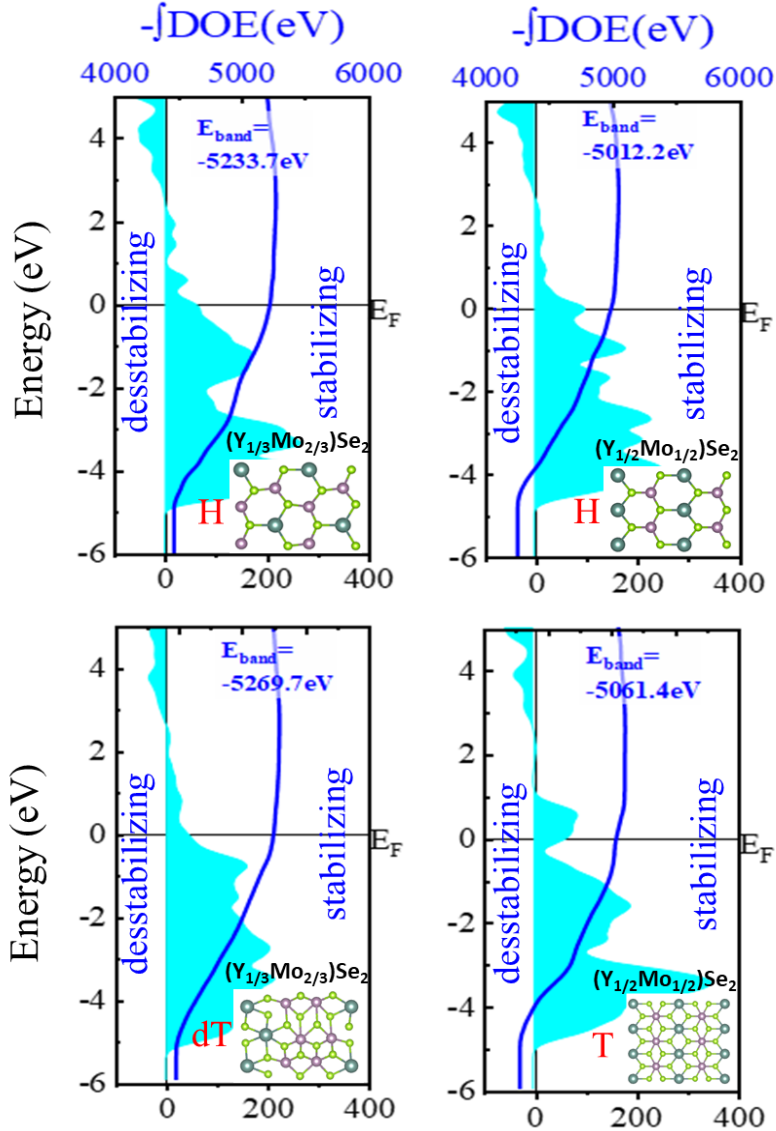

FIG. 6. $DOE$ analysis of the bonding for the cases with H- to dT/T-phase transitions. The $DOE$ energy integrals are shown as blue lines and the total band energy, $E_{band}$, is also given.

# IV. CONCLUSIONS

In conclusion, we performed systematic DFT calculations to obtain a unified understanding to the electron-and hole-driven H-to-dT/T phase transitions in group VI-B monolayer TMDs. Hole injection gives the group VI-B TMDs a phase transition from H to T while electron injection leads a phase transition from H to dT. The origin can be attributed simply to the smaller work function of the T-phase than that of the dT-phase. Our work function analysis can distinguish T and dT phases quantitatively and can give a unified explanation for the phase transitions for both electron doping and hole doping, and is not limited to TMDs. Furthermore, we show that the hole-driven phase transition mechanism can be used to explain the recent experimental observations. Our current work provides a unified understanding of the H-to-T phase transition by hole doping and H-to-dT phase transition by electron doping, and hence helps to expand the possible stable phases of 2D materials.

## ACKNOWLEDGMENTS

This work was supported by the Natural Science Foundation of Hebei Province of China (Grants No. A2021201001, A2021201008), the Natural Science Foundation of China (Grants No. 11904154, 12104124, 51772297), the Advanced Talents Incubation Program of the Hebei University (Grants No. 521000981390, 521000981394, 521000981395, 521000981423), and the high-performance computing center of Hebei University.